\documentclass[prl,twocolumn,amsmath,amssymb]{revtex4}
\usepackage[hiresbb,final]{graphics}

\begin{document}
\title{Droplet traffic in microfluidic networks:\\
A simple model for understanding and designing}
\author{Michael Schindler}
\author{Armand Ajdari}
\affiliation{Laboratoire PCT, UMR ``Gulliver'' CNRS-ESPCI 7083, 10 rue Vauquelin, 75231 Paris cedex 05}

\begin{abstract}
We propose a simple model to analyze the traffic of droplets in microfluidic
``dual networks''. Such functional networks which consist of two types of
channels, namely those accessible or forbidden to droplets, often display a
complex behavior characteristic of dynamical systems. By focusing on three
recently proposed configurations, we offer an explanation for their remarkable
behavior. Additionally, the model allows us to predict the behavior in different
parameter regimes. A verification will clarify fundamental issues, such as the
network symmetry, the role of the driving conditions, and of the occurrence of
reversible behavior. The model lends itself to a fast numerical implementation,
thus can help designing devices, identifying parameter windows where the
behavior is sufficiently robust for a devices to be practically useful, and
exploring new functionalities.
\end{abstract}

\maketitle

Injecting immiscible fluids into a microfluidic system has been found to permit
the controlled generation of droplets or bubbles~\cite{digital}. Rapidly,
\emph{droplet microfluidics} has developed beyond early observations because of
its promises for a diversity of fields~\cite{reviews,GunJen06}: studies of
chemical kinetics and time-controlled synthesis~\cite{kinetics}, high-throughput
screening~\cite{ZheIsm05}, fabrication of remarkable colloidal
objects~\cite{colloids}, combinatorial chemistry~\cite{kinetics}, and even
non-electronic coding or computing functions \cite{PraGer07,FueGarWhi07}. These
tasks require mastering the generation of droplets, their traffic in potentially
complex networks, and sometimes their splitting or merging. One can either
employ passive means by relying on hydrodynamics and capillarity
alone~\cite{digital,reviews,kinetics,colloids,PraGer07}, or integrate active
actuators (electric, dielectric, pneumatic) in the devices~\cite{reviews,Herve}.
Difficulties often arise even in simple geometries if the complex behavior of
dynamical systems (period doubling, chaotic behavior, etc.)
occurs~\cite{Herve,GarFueWhi05}.

\begin{figure}[tb]%
  \centering%
  \includegraphics{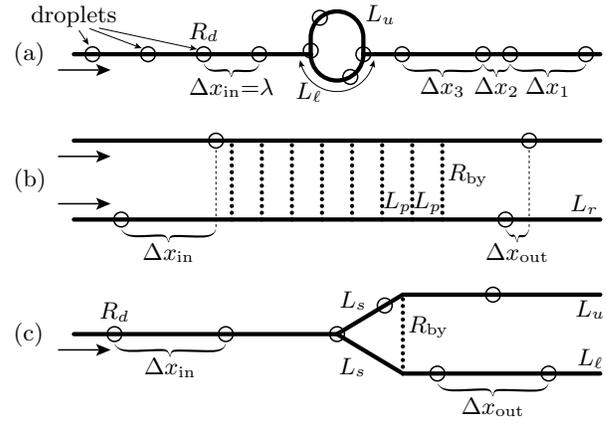}%
  \caption{Graph representations of three recently proposed dual
  networks: (a)~a loop~\cite{PraGer07}; (b) a ladder~\cite{FueGarWhi07}; (c) a
  bypassed T-junction~\cite{GalderAPL}. Full and dotted lines
  correspond to transport channels (accessible to droplets) and ``bypass''
  channels (forbidden to droplets). Droplets are indicated by circles.}%
  \label{fig:devices}%
\end{figure}

We here focus on the traffic of droplets only, without splitting or merging, in
what we name \emph{dual networks}, i.\,e.~networks combining channels that can
be accessed by droplets (\emph{transport channels}) and others that cannot
(\emph{``bypass'' channels}), see Fig.~\ref{fig:devices}. This family of
networks encompasses many recently proposed
solutions~\cite{PraGer07,FueGarWhi07,EnglPRL,GalderAPL}. Building on recent
works \cite{EnglPRL,Jousse}, we introduce a simple numerical tool to analyze the
collective motion of droplets in such a dual network. We show this tool to be
powerful in both illuminating fundamentals ruling the traffic (including
considerations of symmetry and reversibility) and as a means to predict the
capability and robustness of a given design in performing a given function. In
particular, we illustrate these ideas by analyzing three recently proposed
devices as displayed in Fig.~\ref{fig:devices}: a~reversible encoding
loop~\cite{FueGarWhi07} (\ref{fig:devices}a), a ladder for synchronizing
droplets between parallel channels~\cite{PraGer07} (\ref{fig:devices}b), and a
bypassed T-junction for symmetric splitting of a droplet train~\cite{GalderAPL}
(\ref{fig:devices}c).

{\em Simple Model -- }
We model both the flow of the continuous phase and the traffic of droplets in an
arbitrary \emph{dual network} of transport channels and bypasses, represented as
a simple directed graph of nodes and links as in Fig.~\ref{fig:devices}. The
instantaneous flow state is given by a pressure~$P_\alpha$ at each node and a
total flowrate~$Q_i$ in each channel. These variables are related by equations
similar to Kirchhoff's rules for electrical circuits. Mass conservation at node
$\alpha$, taking into account a possible external injection
$Q_\text{ext}^{(\alpha)}$, reads
\begin{equation}
  \label{eq:node}
  \sum_\text{$i$ connected to $\alpha$} \epsilon_i Q_i = Q_\text{ext}^{(\alpha)},
\end{equation}
where $\epsilon_i=\pm 1$ account for the orientation of the link. In channel~$i$
the flowrate~$Q_i$ is related to the pressure difference between its two end
nodes $\alpha_{\text{in}(i)}$ and $\alpha_{\text{out}(i)}$,
\begin{equation}
  \label{eq:edge}
\Delta P_i=  P_{\alpha_{\text{in}(i)}} - P_{\alpha_{\text{out}(i)}} = R_i Q_i.
\end{equation}
Equation~\eqref{eq:edge} defines the \emph{hydrodynamic resistance}~$R_i$ of
channel~$i$. For flows without droplets at low Reynolds number it reduces to a
constant value~$\bar R_i$, proportional to the viscosity and to the
channel length $L_i$: $\bar R_i\propto L_i$.

Droplets or bubbles in the channel significantly modify the picture in a quite
complex way \cite{EnglPRL,Jousse,GunJen06}. A simplification occurs if the
droplets are sufficiently distant from one another (typically a few channel
diameters away) so that their flow perturbations do not interact. In the
mentioned experiments the ratio of distance to diameter is approx.~10 (loop and
ladder) and 5 (bypass junction). Each droplet then yields an additive
increment~$R_d$ to the resistance. In this paper, we make the further
simplification that all droplets are equal and so are the cross-sections~$S$ of
the channels transporting droplets (these restrictions can be easily lifted in
the numerical model), so that for $n_i$~droplets in channel~$i$,
\begin{equation}
  \label{eq:dresist}
  R_i = \bar R_i + n_i R_d,
\end{equation}
Identical droplets with no interaction move all at the same velocity~$v_i$ in a
channel with a given flow rate~$Q_i$, which we write with a
proportionality factor~$\beta / S$,
\begin{equation}
  \label{eq:dveloc}
  v_i = \beta Q_i / S.
\end{equation}
In general, both $R_d$ and $\beta$ will vary with~$Q_i$
\cite{Bretherton61,Jousse}. The simplifications in \eqref{eq:dresist} and
\eqref{eq:dveloc}, however, have proved to hold in experimental
realizations~\cite{EnglPRL}.

The flow state, consisting of all pressure and flowrate values, is uniquely
determined by the set of linear equations \eqref{eq:node} and \eqref{eq:edge}
when complemented by appropriate boundary conditions at the end-nodes,
i.\,e.~specification of either the pressure or the externally injected flowrate.
An important point is that this flow state persists until a droplet arrives at a
node and enters another channel, thereby modifying the set of resistances. To
describe droplet traffic, we thus track the locus of the droplets (modeled as
points) along the channels. When a droplet arrives at a node where there are
several channels to choose from, a \emph{selection rule} is required. We focus
here on locally symmetric T-junctions for which the droplets enter the channel
with the instantaneous larger flow rate~\cite{EnglPRL,Jousse}.

Altogether, we propose the following algorithm: (i)~Compute the pressures and
flowrates for a given set of droplet positions using Eqs.~\eqref{eq:node} and
\eqref{eq:edge}; (ii)~determine the droplet velocities according to
Eq.~\eqref{eq:dveloc}; (iii)~determine the next time when either a droplet
arrives at a junction or a new one is injected, and advance all droplets to this
time; (iv)~decide which route the triggering droplet takes, update the channel
resistances of the affected channels, and return to step~(i).

Clearly, step~(iii) is the big time saver that allows fast computations of rather
involved histories. This scheme can be straightforwardly adapted to networks
with channels of different cross-sections and to other junction geometries, but
the present version is sufficient to revisit recently proposed devices and
discuss fundamental issues.

{\em Reversibility -- }
\begin{figure}[b]%
  \centering%
  \includegraphics{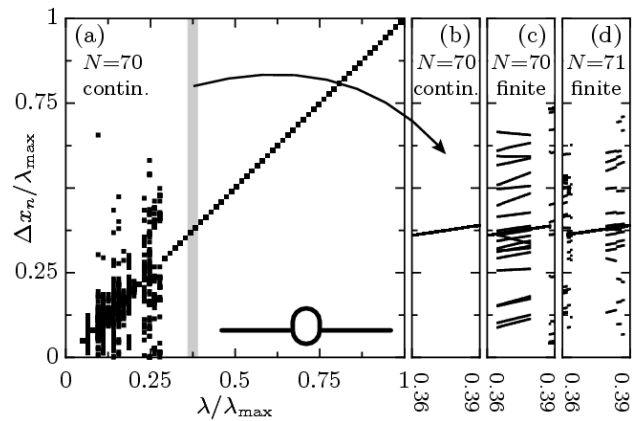}%
  \caption{Final distances~$\Delta x_n$ between 70~droplets after a return
  journey through the loop of the device in Fig.~\ref{fig:devices}a, as a
  function of the initial period~$\lambda$. Reversibility is observed where all
  69~symbols collapse on the diagonal, see also the zoomed panel~b. The results differ for a finite droplet
  train (panels c and d), where also the number of droplets matters. The
  parameters $L_u/L_\ell=1.112$, $R_d/\bar R_\ell=1.5$, are compatible with data
  from Ref.~\cite{FueGarWhi07} \cite{period7}.}%
  \label{fig:loop}%
\end{figure}%
\begin{figure}[b]%
  \centering%
  \includegraphics{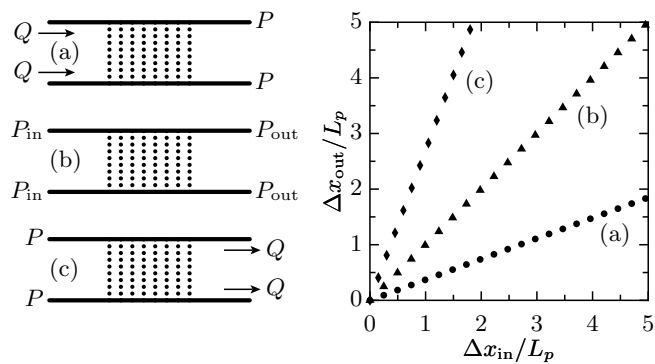}%
  \caption{Change of droplet misalignment in the ladder device. Only the
  boundary condition (a) leads to a contraction; fixed pressures on both ends
  (b) do not affect the offset, while it grows in (c). $R_d{/}\bar R_p {=}
  20$, $R_\text{by}{/}\bar R_p {=} 5$, $L_r{/}L_p {=} 5$.}%
  \label{fig:ladderdata}%
\end{figure}%
Fuerstman~\textit{et\,al.}~\cite{FueGarWhi07} presented a device that allows
reversible ``encoding'' of information in interdroplet distances: Two long
channels, used to store droplet series, are connected by a slightly asymmetric
loop (Fig.~\ref{fig:devices}a). A regular train of droplets, injected through
one channel into the loop, exits the loop as a complex modulated sequence. The
interdroplet distances result from the ``decisions'' taken by the droplets when
entering either branch of the loop. Each decision is affected by the presence of
previous droplets in the loop. The slight asymmetry of the loop gives rise to
periods of $2, 3, \ldots 7$ in the outgoing sequence, depending on the initial
interdroplet distance~\cite{FueGarWhi07}. When we apply our simple model using
parameters reasonably related to those of the experimental realization, we do
obtain a comparable cascade of periods ($2, 3, 5, 7, 9$ etc.), as a
characteristics of dynamical systems at work in this network~\cite{period7}.

A remarkable result by Fuerstman~\textit{et\,al.} \cite{FueGarWhi07} is that
upon reversal of the flow direction, the starting sequence of droplets is
restored, i.\,e.~the system exhibits \emph{reversibility}. This is far from
being obvious, keeping in mind that the selection rule is not intrinsically
reversible.

To investigate the question of reversibility further, in our numerics we send a
periodic train of droplets through the loop until $N$~droplets have reached the
other side, then invert the flow direction and send them back. The resulting
$(N\,{-}\,1)$~interdroplet distances after this return trip through the loop are plotted in
Fig.~\ref{fig:loop}a for several values of the initial distance~$\lambda$,
ranging from $0$ to the maximum for which two droplets interact,
$\lambda_\text{max} = L_\ell\bigl[1 + (\bar R_\ell + R_d)/\bar R_u\bigr]$ (see
Fig.~\ref{fig:devices}a for the parameters). The most striking feature of
Fig.~\ref{fig:loop}a is that it predicts windows of values of~$\lambda$ for
which the system behaves reversibly and other windows for which it does not
(points not on the diagonal). It also sheds light on the importance of the
precise conditions at flow inversion: For the same parameter
interval~\cite{period7}, both full reversible behavior (Fig.~\ref{fig:loop}b)
and a complicated interrupted pattern (Fig.~\ref{fig:loop}c) can be produced. In
panels \ref{fig:loop}a and \ref{fig:loop}b the droplets were perpetually
injected, such that the loop was not empty when changing the flow direction.
Conversely, in Fig.~\ref{fig:loop}c and \ref{fig:loop}d, a finite train of
$N$~droplets was sent to and fro, with an empty loop in the meantime. This
latter case is hardly reversible and depends on the precise number of droplets.

Let us underline that non-reversibility is deeply linked to the selection rule,
and thus to the presence of junctions between transport channels. In a dual
network with no such junctions, one expects the behavior to be reversible.

{\em Reversibility and Symmetry of the Design -- }
This observation leads us to a rather generic point regarding the kind of
functions that a fore/aft symmetric device can achieve if the dynamics is
reversible. A good support for this discussion is the recently
proposed~\cite{PraGer07} symmetric ladder network of Fig.~\ref{fig:devices}b.
Two transport channels are connected by a few (narrower) bypass channels. The
device has been shown to perform the following function: Two droplets arriving
with a given delay in the entrance channels leave the system with a much reduced
distance, allowing synchronization between the two channels. Why is this
remarkable? On one hand, (A)~as the device is symmetric in design, if distances
are reduced going one way, they should also be reduced going the reverse way. On
the other hand, (B)~the absence of junctions leading to ``decision making
steps'' implies reversible dynamics, so that if the distance decreases one way,
it should increase the other way: Such paradoxical statements are commonly used
in low Reynolds number fluid dynamics to show that there can be no net effect.
Here, experiments tell us otherwise.

Running our model immediately resolves the paradox in a manner that is
practically relevant: The function can be achieved only if the boundary
conditions are not symmetric! For example, if we apply constant flow rates in
the two entering arms and a fixed pressure at the two exiting nodes, we recover
qualitatively the phenomenology described in Ref.~\cite{PraGer07} with distance
reduction (see case a in Fig.~\ref{fig:ladderdata}). This reduction is not
forbidden by the symmetry argument, as changing the sign of the flowrates and of
the pressures yields a pattern of boundary conditions that is not the symmetric
image of the initial situation, eliminating argument~(A). In this new situation
(case c in Fig.~\ref{fig:ladderdata}) the ladder acts as a distance expander
rather than a distance contractor, in accordance with~(B). Conversely, if we use
fixed pressures at all end nodes, then the effect vanishes, as required by the
symmetry argument (A+B) above (case b in Fig.~\ref{fig:ladderdata}). Our
analysis thus clarifies the limits of use of this device as a synchronizing
functional block in a network context.

At this stage, we have demonstrated that our model reproduces qualitatively and
quantitatively the performance of existing devices. It illuminates the essential
role of the imposed driving conditions~\cite{loopremark}, and it provides a
priori indication as to the dynamic behavior (period doubling,
reversibility,~etc.) as a function of design parameters (geometrical symmetry)
and operational parameters (droplets distance, etc.).

{\em Designing Devices: Optimization and Robustness -- }
\begin{figure}[tb]%
  \centering%
  \includegraphics{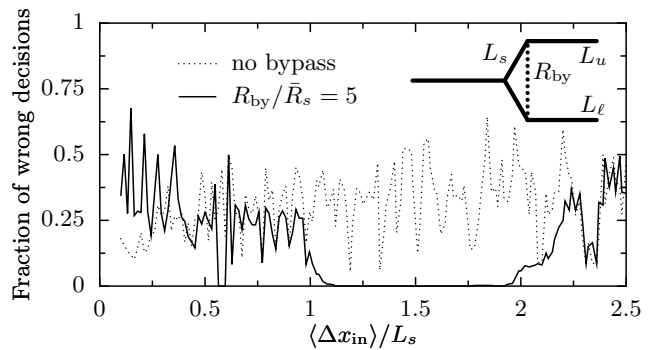}%
  \caption{Fraction of wrong (not strictly alternating) decisions in T-junctions
  with and without bypass (Fig.~\ref{fig:devices}c). The interval without errors
  around the average droplet distance $\langle\Delta x_\text{in}\rangle \approx
  1.5 L_s$ corresponds to the experimental findings in Fig.~3 of
  Ref.~\cite{GalderAPL}. The robustness of the behavior is challenged by an
  asymmetry ($L_u/L_\ell = 1.01$) and a random perturbation of the incoming
  distances (standard deviation is $5\%$ of the average value). $R_d/\bar R_s =
  0.3$.}%
  \label{fig:bypassdata1}%
\end{figure}%
The speed of our numerical scheme permits fast optimization of identified
designs. For example, we can scan the effect of different parameters of the
ladder device (Fig.~\ref{fig:ladderdata}a), such as the number of bypasses,
their resistance, their distance, or the resistance of the droplets. The
parameter $R_d/\bar R_p$ has the strongest influence: the closer the bypasses,
the better. Conversely, making the side arms of the transport channels ($L_r$)
very long, has a negative effect on the contraction. Increasing the number of
bypasses from 3~to 10, or making them 100~times less resistant only weakly
improves the contraction result.

Beyond this optimization in an ideal world, an essential point for suggested
designs to turn into efficient working devices is their robustness to both
fabrication errors and operational variations---virtues that are rarely
quantified. We demonstrate the easy use of our tool to qualify the bypassed
T-junction of Fig.~\ref{fig:devices}c, proposed in Ref.~\cite{GalderAPL}. The
aim of this engineered junction is to provide reliable alternation of the
channel chosen by incoming droplets. How it works is detailed in
Ref.~\cite{GalderAPL} -- in short, the bypass channel imposes similar pressures
at both its ends. A droplet having chosen one route is then able to ``block''
its channel such that the next droplet favors the other one due to its larger
flowrate. With our numerical scheme we immediately recover the same parameter
window for which the experimental design has been shown to be successful (cf.\
Fig.~3\ in Ref.~\cite{GalderAPL}).

We can systematically explore the robustness of the alternating behavior to
imperfections of the device. These may consist of a ``quenched'' design
asymmetry, such as a small systematic bias in the decision rule or a difference
in arm lengths; or they may come as random fluctuations in the incoming droplet
distance or in the pressures at the outlets. Figure~\ref{fig:bypassdata1} shows
the effect of a random Gaussian perturbation of the incoming droplet distances,
combined with a slight asymmetry of the channel lengths behind the bypass. Both
perturbations lead to the complete failure of an unbypassed junction, whereas
even a thin bypass robustly yields a perfect alternation in a significant window
of parameters.

Playing with this analysis we find another remarkable property of the bypassed
T-junction, namely noise reduction in the droplet distances. The distances in
the two outgoing sequences have a smaller standard deviation than the incoming
train (data not shown). This effect operates for a single junction and can be
amplified by using a cascade of several junctions, connected in such a way that
the average hydrodynamic resistances are balanced at each stage of the
cascade~\cite{patent}. At each level of the cascade the noise is reduced because
the relative standard deviation of the distance between over-next droplets is
only $1/\sqrt{2}$ times that between adjacent droplets.

{\em Conclusion -- }
We have proposed a simple yet efficient fast numerical tool to analyze the
traffic of droplets in ``dual microfluidic'' networks. Through an analysis of
three recently proposed devices, we have shown the power of this tool for
understanding fundamental issues, such as the role of the boundary conditions,
of geometric symmetry and of the occurrence of reversible behavior. We have left
aside physical effects, in particular capillarity, that could add further
functionalities, such as trapping in tapered channels, splitting and merging of
droplets at junctions, etc. However, we believe, that the presented method will
stimulate further experiments to explore the validity of its central ideas.
After such a validation step it will provide a guide for the design of devices,
and for exploration of new functionalities. In particular, it lends its ideas to
\emph{robust} passive solutions that are likely steps for droplet (or bubble)
microfluidics to practically come up to the hopes it has raised in many fields.

This work was supported by the French National Research Agency (ANR) via project
\emph{Scan2}, and we thank all members of this project for stimulating
discussions.



\begin{thebibliography}{99}
\bibitem{loopremark}
  In the special case of a single pair of end-nodes (e.g.~the loop) the boundary
  conditions do not affect the sequence of events but only the pace at which
  they occur.
\end{thebibliography}
\end{document}